# The Development of Epitaxial Graphene For 21st Century Electronics.

Walt A. de Heer

Georgia Institute of Technology

## Abstract

Graphene has been known for a long time but only recently has its potential for electronics been recognized. Its history is recalled starting from early graphene studies. A critical insight in June, 2001 brought to light that graphene could be used for electronics. This was followed by a series of proposals and measurements. The Georgia Institute of Technology (GIT) graphene electronics research project was first funded, by Intel in 2003, and later by the NSF in 2004 and the Keck foundation in 2008. The GIT group selected epitaxial graphene as the most viable route for graphene based electronics and their seminal paper on transport and structural measurements of epitaxial graphene was published in 2004. Subsequently, the field rapidly developed and multilayer graphene was discovered at GIT. This material consists of many graphene layers but it is not graphite: each layer has the electronic structure of graphene. Currently the field has developed to the point where epitaxial graphene based electronics may be realized in the not too distant future.

## Introduction: a brief history of graphene

Graphene has a long history. Already in the 1800's it was known that graphite was a layered material that could be exfoliated by various methods. The famous chemist Acheson[1] had developed exfoliation methods (that he called deflocculation) to produce colloidal suspensions of small graphitic flakes that he called "dags". These colloidal graphite suspensions have been extensively used in the electronics industry as a conducting paint to produce conducting surfaces in vacuum tubes. In 1907 the material was described as follows: "The graphite is in what Mr. Acheson calls the "deflocculated" condition, a condition of fineness beyond that attainable by mechanical means, a condition approaching the molecular slate." Indeed, these suspensions probably do contain graphene flakes as Acheson suspected. He then found that coating metals with deflocculated graphite prevented corrosion.

But Acheson also discovered another material. He produced silicon carbide (SiC) that he called carborundum. It was first used as an abrasive. He then found that upon heating silicon carbide, silicon evaporated leaving behind very pure graphite. Exfoliated graphite and graphite on silicon carbide would become extremely important a century later for their electronic properties.

In 1859, using nitric acid and potassium chromate, Brodie produced "graphite oxide" consisting of graphene sheets with epoxide groups bridging adjacent carbon atoms. Immersing this material in water causes the individual layers to separate.

In 1962, H-P Boehm[2] recognized that graphite oxide monolayers could be reduced to graphene by reducing it with hydrazine. He next was able to measure the thicknesses of these graphene layers using a combination of surface measurements

X-ray scattering and electron microscopy (Fig. 1). Boehm thereby was the first to produce and to measure freely suspended graphene flakes. In 1982 Boehm officially coined the name "graphene" which he saw as a macromolecule (the "ene" ending refers to hydrogen at the edges, as for example in benzene)[3].

The importance of silicon carbide (a wide bandgap semiconductor) for electronics had been known since the early 1900's, when SiC crystals were used as diode detectors for radio receivers. High quality SiC crystals are now grown using the Lely process.

In 1975 van Bommel et al[4] studied the graphitic layers that are formed when hexagonal silicon carbide is heated in vacuum (Fig. 1). They confirmed Acheson's observation that the material graphitizes because silicon sublimes from the crystal surfaces. However they also noted that the different crystal faces graphitized differently. Graphite grown on the silicon terminated face (Si-face) was clearly epitaxial (that is, aligned with the silicon substrate) while the carbon terminated face (C-face) seemed to be rotationally disordered. Further studies by Forbeaux et al.[5] in 1998 confirmed these results, but also indicated that the graphite layers were essentially decoupled from the substrate, causing them to conclude, that "layer by layer growth opens up the possibility of an isolated single graphene layer "floating" above the substrate ".

Graphene had also been produced by other methods, including by chemical vapor deposition (CVD) of carbon containing gasses on various metals. The two dimensional crystal properties were well known and intensely studied in the surface science community (for a review, see Gall[6]).

In 1999 Ruoff and coworkers[7] recognized that graphene might be important for various reasons, and developed a mechanical method to exfoliate graphite. They etched vertical pillars in a graphite crystal and demonstrated that very thin graphite flakes could be mechanically slid off of the pillars (see Fig. 3a). Their goal was to produce graphene sheets by this method.

The electronic properties of graphene were well understood since 1947, when Wallace[8] used simple tight-binding calculations to calculate its band structure (Fig. 1). In 1957 McClure[9] calculated the properties of graphene in magnetic fields and demonstrated the unusual magnetic field dependence of the Landau levels. Others confirmed and expanded these calculations. In 1998 Ando[10] recognized the equivalence of Wallace's graphene bandstructure with the dispersion relation of a massless fermion (i.e. a neutrino), in which the spin was replaced by a pseudospin (that actually characterizes the bonding and antibonding character of the wave functions). He also recognized the resulting non-trivial Berry's phase responsible for the unusual quantization of the Landau-levels as revealed in the quantum Hall effect. Moreover, he predicted that the pseudo-spin character would be responsible for the reduction of backscattering both in nanotubes and in graphene.

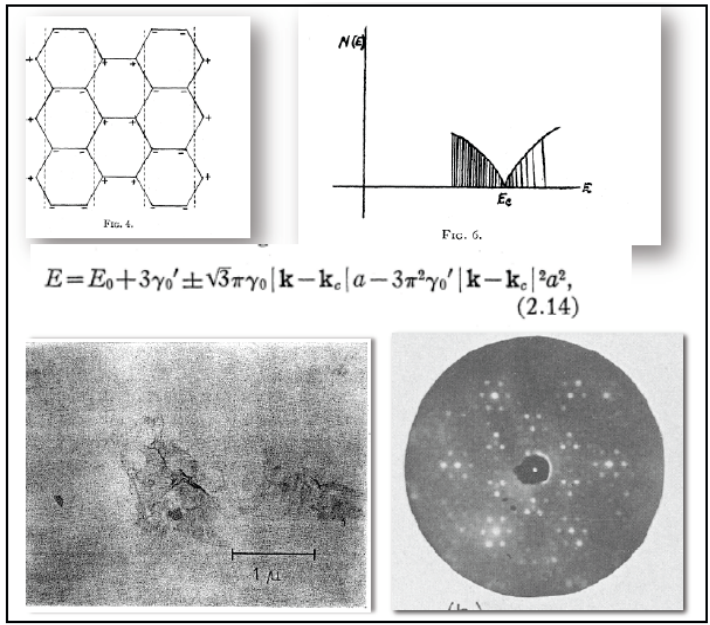

**Fig. 1**
The early days: Top, graphene structure and density of states calculated by Wallace (1947) using the tight binding model[8], which showed revealed the conical band structure (equation, in center, recently renamed the "Dirac cone"). Bottom left, electron micrograph of graphene flake electron from Boehm (1962)[2]; Bottom right: LEED image of Si-face epitaxial graphene, by van Bommel (1972)[4].

**The carbon nanotube connection**

The ultimate demise of silicon electronics as the primary driving force of electronics innovation had been predicted for decades. This "end of Moore's law" dilemma spurred research around the globe in the search for a viable alternative material. In the 1990's, carbon nanotubes had been discovered and they were first seen primarily as interesting for their great mechanical strength: unusual proposed early applications included elevators to space.

Iijima[11] had pointed out that carbon nanotubes could be seen as rolled up graphene sheets, and Mintmire and White[12] observed that depending on their helicities, carbon nanotubes could be either metals or semiconductors. This very interesting observation suggested that nanotubes might be used for electronics. In 1998, my group demonstrated that carbon nanotubes are ballistic conductors[13]. This implied that electronics move large distances (order of microns) without scattering, which is very favorable for electronics. Independently and contemporaneously, Todorov and White[14] provided a theoretical explanation for ballistic transport in carbon nanotubes.

That same year Tans et al[15] demonstrated the first nanotube transistor. It consisted of a nanotube with metal contacts on either end that was placed on an oxidized degenerately doped silicon wafer. This geometry allowed the nanotube to be "back gated": applying a voltage to the silicon wafer induces charges on the nanotube, thereby turning it on and off. Nanotube transistors were remarkable, and over time they did indeed demonstrate impressive characteristics. However there remained a fatal flaw: while the nanotube resistance can be very low, for quantum mechanical reasons, the contact resistances are very high and of the order of 10 k Ohm! Heat is dissipated at the contacts making them extremely fragile. Besides the contact and interconnect problem, there are also problems with nanotube placement. Hence, it does not appear likely that nanotube electronics will be a viable successor for silicon.

But a valuable lesson had been learned: carbon might play a role in future electronics, setting the stage for the next development.

**The birth of graphene electronics.**

In early June, 2001, a colleague called me and asked for a carbon nanotube tight binding program that I had written years earlier. I had lost the program and stated to rewrite in on June 9. It is very easy to write the program for finite ribbons or nanotubes. It involves a simple matrix with the rows and columns representing the carbon atoms. In the matrix, ones are placed whenever two carbon atoms are nearest neighbors and zeroes otherwise. A nanotube is simply a ribbon for which the two edges are connected. I first ran the program for a ribbon, and to my amazement, the resulting density of states resembled that of a nanotube, both with semiconducting and metallic varieties (Fig.2). A full bandstructure calculation confirmed this (Dresselhaus had done similar calculations[16]).

It was therefore clear that graphene ribbons were electronically similar to nanotubes, and I realized that this was extremely important. For the next month I expanded this idea and calculated the electronic properties of a great many structures, including rough edges, junctions, magnetic fields and many other features. I developed a program that would allow any graphene structure to be drawn after which the electronic structure would be calculated. It turns out that a "brick wall" topology (where the mortar lines represent bonds) is identical to that of graphene. This greatly simplifies the indexing of the atoms.

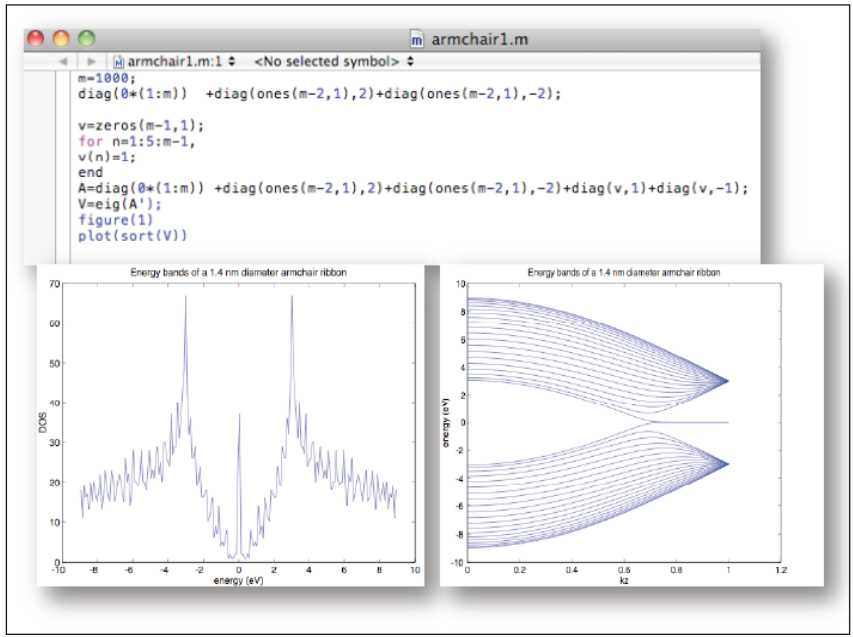

*Fig. 2*
Tight binding model. Top. Simple finite graphene ribbon tight binding program, June 9, 2001. Bottom left. Ribbon density of states, right, ribbon bandstructure. Note that at the time I used the standard nanotube nomenclature, so that armchair and zigzag interchanged.

I also developed a program to propagate waves through finite graphene structures, which was easy since the wavefunctions could be exactly calculated. These calculations convinced me that graphene based electronics was possible, and that the major hurdles in carbon nanotube based electronics were potentially resolved. Specifically, wide ribbons (functioning as the metal leads) could be connected to a narrow ribbon (functioning as a semiconducting channel) to produce graphene transistors. Expanding on this theme, entire integrated structures could be developed simply by appropriately cutting a graphene sheet.

In July 2001, at the NT01 nanotube conference in Potsdam I discussed graphene based electronics with Robert Haddon, who was immediately very interested, especially because of the connection to his work on fullerenes and nanotubes. He contemplated the intriguing prospects of chemical functionalization of graphene and its edges. This is of course a very important aspect, when one realizes that graphene is actually an organic macromolecule.

The next step was then to assemble a research group to work on this project. Claire Berger and Phillip First were willing to participate. We then needed a viable platform. We became aware of the various graphene production methods mentioned above and started work in three directions, as documented in our proposal to the NSF in 2001 (Fig. 3a,b,c) [17].

The first method described in the 2001 proposal involved an exfoliation method using Ruoff's pillars[7] to deposit graphite flakes on oxidized degenerately doped silicon wafers to make field effect transistors (Fig.3a), very similar to the method used by Novoselov et al[18] in their 2004 paper on the electric field effect of thin graphite flakes (the Scotch tape method was not used at this time). We anticipated that this method would only be useful for proof of principle demonstrations. The second method involved the graphitization of silicon carbide. At the time the proposal was submitted Phil First had graphitized a few silicon carbide samples in his UHV chamber to provide preliminary results (Fig. 3c). The third method involved CVD growth on metals.

In the 2001 proposal[17] listed the following five important properties of graphene based electronics:

*(i)* size-tunable electronic bandgaps,

*(ii)* chemical robustness,

*(iii)* immunity to electromigration (a major problem in nanoelectronics),

*(iv)* high current capability, and

*(v)* electrically tunable conductivity using the field effect via a proximal gate electrode ("gate-doping");

and concluded with the statement:

*Our vision is nothing less than a new form of large-scale integrated electronics based on ultrathin films of lithographically-patterned graphite.*

Items ii, iii and iv all related to the extreme stability of graphene, which indeed is the most stable macromolecule known.

This proposal was not funded, as were 3 others that followed in 2002. Referees generally thought that graphene based electronics were of little technological or scientific value. Graphene, had been around for a long time and it was mainly a nuisance, a poison for catalysts. From their reports it was abundantly clear that only a demonstration would convince them.

In late 2002, we had the necessary demonstrations. We had successfully produced epitaxial graphene using van Bommel's method and patterned it using e-beam lithography methods to make simple Hall bars and transport measurements, demonstrating the 2D gas properties of epitaxial graphene. We had also demonstrated top gating. Most of the transport measurements at that time were conducted at the mesoscopic science laboratories of the Neel institute of the CNRS in Grenoble, France, which was and still is Claire Berger's base institute.

These data were compiled in a proposal to the NSF in June 2003 (Fig.3 d,e,f)[17]. At that time, the mobilities were not good, typically on the order of 15 cm$^2$V$^{-1}$s$^{-1}$ (Fig. 5b) and it was clear why. Atomic force microscopy studies showed that the graphitized surfaces were pitted, resembling Swiss cheese (Fig. 5a). Nevertheless, we contacted Intel Corp. who were very interested in our work and provided our first significant funding in September 2003. In June 2003 we also wrote our first patent, that claimed the invention of graphene based electronics, incorporating the concepts we had developed in the previous two years. The patent, signed by de Heer, Berger and First, was issued in 2006[17].

The poor mobilities were of great concern and clearly linked to the production method. I started experimenting with alternative furnace methods and found that by confining the silicon carbide samples in a graphite cylinder, the quality of the graphene dramatically improved. Mobilities exceeding 1000 cm$^2$V$^{-1}$s$^{-1}$ were commonly achieved and the AFM images revealed that the surfaces were now not pitted (Fig. 5 c). In summer 2004 we produced and measured a Hall bar on the Si face with a mobility of about 1100 cm$^2$V$^{-1}$s$^{-1}$ that exhibited large Shubnikov de Haas oscillations and signatures of the quantum Hall effect (Fig. 5d). At the time it was difficult to measure the thickness of the layers accurately and we relied on Auger electron spectroscopy. We were not aware that this method overestimates the graphene thickness, because it did not take into account the electrically inert carbon rich buffer layer on the Si-face. Therefore we thought that that we were measuring an ultrathin graphite trilayer while in fact it was a graphene monolayer. These measurements thereby were the first graphene transport measurements as we pointed out later.

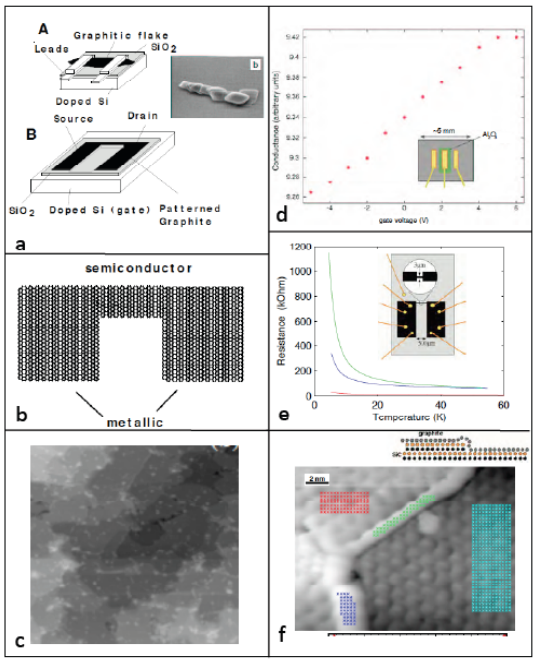

*Fig. 3*
Development of graphene electronics in early proposals a,b,c 2001, d,e,f 2003.[17] (a) Proposed field effect transistor composed of a thin graphite or graphene flake deposited on an oxidized degenerately doped silicon wafer. The flakes were to be produced using the Ruoff pillar method[7] (inset). (b) Proposed graphene transistor with metallic graphene leads and a semiconducting graphene channel. (c) Early example of epitaxial graphene on the Si face of a SiC chip. (d) Field effect an and unpatterned graphene sheet. (e) Patterned graphene ribbon resistance versus temperature. (f) STM image of epitaxial graphene. Also reported were STS results demonstrating that the graphene layer was continuous over the substrate steps. These data were published in 2004[19].

We published these results in Dec. 2004 in our paper titled "Ultrathin epitaxial graphite: 2D electron gas properties and a route toward graphene-based electronics"[19]. We presented in detail our vision for graphene based electronics and we showed extensive structural and electronic characterization of virgin and patterned samples. We had already presented most of these results in the March meeting of the APS earlier that year[20], in a session that was chaired by Millie Dresselhaus. Phillip Kim also presented graphite flake transport results at that meeting.

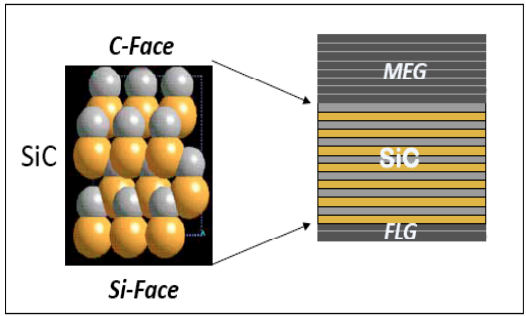

*Fig. 4*
Schematic diagram of graphitized silicon carbide, showing the atomic structure (dark atoms are carbon, light atoms are silicon). Multilayer epitaxial graphene (MEG) grows on the C face (with up to 100 decoupled graphene layers) and few layer graphite (FLG) grows on the Si face (with up to about 10 Bernal stacked graphene layers).

**Beyond the monolayer: the discovery of multilayer epitaxial graphene.**

The explicit mission of our group is to explore the possibility of graphene based electronics. It recognizes that the properties of graphene and its unique bandstructure make it an ideal candidate material for post CMOS electronics.

Epitaxial graphene monolayers are clearly very important and interesting. We demonstrated the quantum Hall effect in high mobility C-face monolayers[21] (Fig. 6a). However, our measurements demonstrated that multilayers could be even more important, especially for applications (Fig. 4). In 2005 we were developing methods to pattern Hall bars on both the Si and the C face of graphene, using primarily e-beam lithography methods. We noted quite soon that the C-face consistently exhibited higher mobilities than the Si face. Moreover, we found from transport measurements that the C face samples *always* exhibited the signature of graphene, as was clear from the non-trivial Berry's phase in the Landau levels[22]. In our 2006 Science paper we explicitly demonstrated a multilayer graphene ribbon grown on the C-face had moblities exceeding 30,000 $cm^2V^{-1}s^{-1}$ and evidence for quantum confinement[22]. It specifically did not show the typical magnetoresistance signature of thin graphite (cf Novoselov[18]).

At the same time, our collaborators in Grenoble were measuring the infrared absorption properties of C-face graphene multilayers in magnetic fields[23]. They found that the Landau levels dispersed as they would for free graphene layers. This meant that this material is not at all like graphite, but that the layers are essentially decoupled from each other. Moreover, the moblities of the material were found to be extremely large (order of 250,000 cm$^2$V$^{-1}$s$^{-1}$)[24]. All of these measurements indicated that we had found a new and important graphene based material that we named multilayer epitaxial graphene (MEG).

Phil First had demonstrated moiré patterns due to the interference of two rotated layers (Fig.6b). Ed Conrad had performed many crystallographic studies (cf Fig. 6 c) of MEG and confirmed the "rotational stacking" that is, the non-Bernal stacking of the layers[25][26]. He recognized that the stacking involved specific commensurate graphene structures. Subsequently the electronic decoupling of the layers was explained theoretically[25].

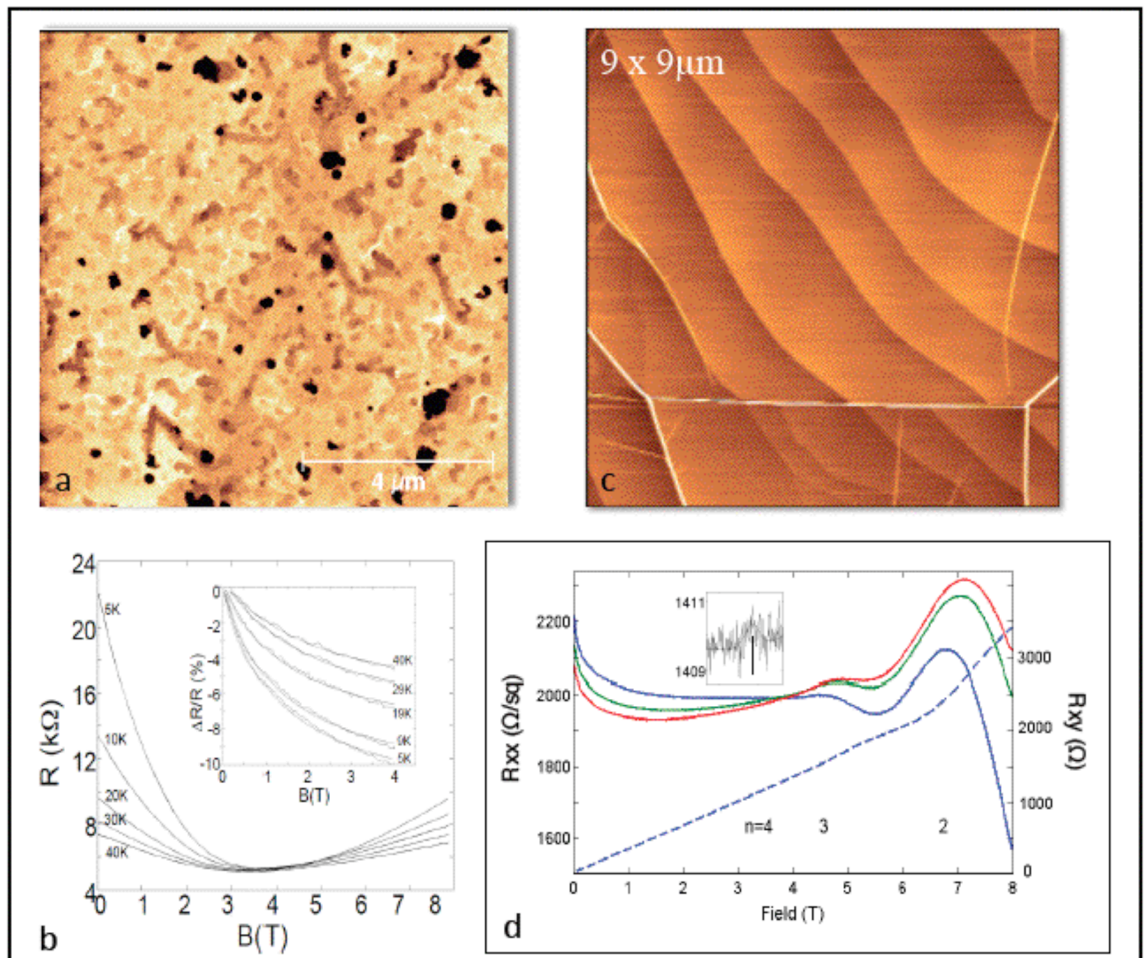

*Fig. 5*
Examples of epitaxial graphene and transport properties. (a) AFM image of typical Swiss cheese morphology of graphene grown by the van Bommel method (in UHV). (b) Conductance (µS) versus magnetic field for UHV grown graphene Hall bar, the mobilities is 15 cm$^2$V$^{-1}$s$^{-1}$ [19].; (c) AFM of typical furnace grown MEG showing substrate steps and pleats on the graphene (note the absence of obvious defects). (d) Transport measurement of monolayer epitaxial graphene Hall bar on the Si face, showing precursor behavior to the quantum Hall effect. Mobility is 1,100 cm$^2$V$^{-1}$s$^{-1}$ [19].

It is relevant to note that at that time many theoretical physicists working on graphene were focused on freestanding monolayers, and they were not interested in dealing with the complications of substrates or multilayers. Many believed that the silicon oxide surface on which exfoliated graphene flakes were deposited did not perturb the electronic structure. However, for metals and SiC they thought that the substrate essentially perturbed the electronic structure of the graphene, even though experimentally this was long known not to be the case. On the other hand, as explained above, exfoliated graphene deposited on oxidized degenerately doped silicon wafers is ideally suited for 2D electron gas studies, because the charge density on the graphene can be continuously varied. This is not so simple in epitaxial graphene.

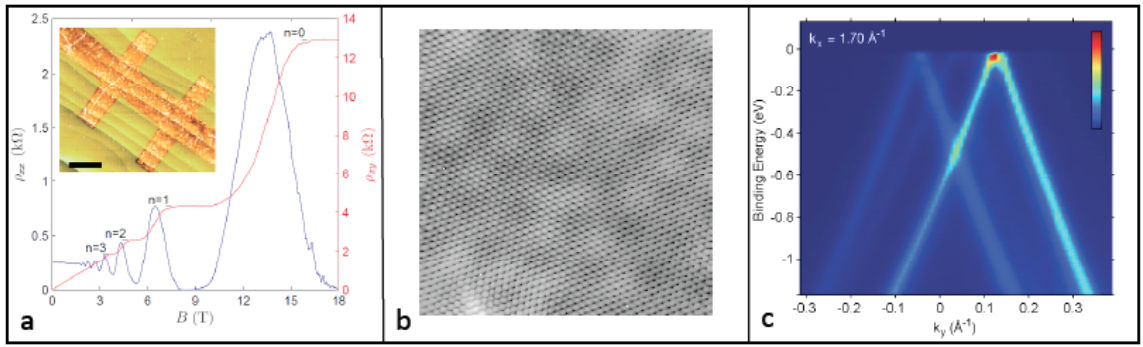

**Fig. 6**
C-face epitaxial graphene. (a) Quantum Hall effect of a graphene monolayer with a mobility of 20,000 cm$^2$V$^{-1}$s$^{-1}$ [21]; (b) STM image of MEG showing the characteristic moiré pattern; (c) ARPES of MEG showing two intersecting Dirac cones due to two adjacent layers[26]. Due to the electronic decoupling of the layers, the cones do not perturb each other.

Many graphene measurements have since been performed on MEG by various collaborators. Notably scanning probe experiments performed in Joe Stroscio's laboratories at NIST using high magnetic fields and low temperatures spectacularly demonstrated properties of this material in the quantum Hall regime, by essentially imaging the electronic wave functions in the graphene[26]. These experiments also showed the extreme high quality of (at least) the top graphene layer of MEG, which was shown to be continuous over the whole surface.

The carbon layers grown on the Si face is distinctly graphitic: it is Bernal stacked and the layers are electronically coupled. While these thin graphite layers are certainly useful, the relatively low mobilities of graphene monolayers appear to be problematic for applications and fundamental physics.

Ted Norris[27] of Michigan State University performed laser pump-probe measurements to demonstrate the terahertz response of MEG. These experiments are relevant, and explore the ultrahigh frequency domain, which is the next important frontier in electronics.

MEG is also ideally suited for chemical functionalization and intercalation studies. At GIT we had first demonstrated that MEG on a SiC chip can be converted into multilayer graphene oxide[28] using standard oxidation processes previously used for graphite. This material is a wide bandgap semiconductor, because the graphitic $sp^2$ bonds are converted into diamond like $sp^3$ bonds. Graphene functionalizaton performed by this method is crude. Recently much more controlled experiments have been performed by Haddon and coworkers to affect a related conversion with aryl groups[29].

These are but a few of the many experiments that are currently in progress on MEG, which may prove to be the ideal material for graphene based electronics. However much work remains to be done.

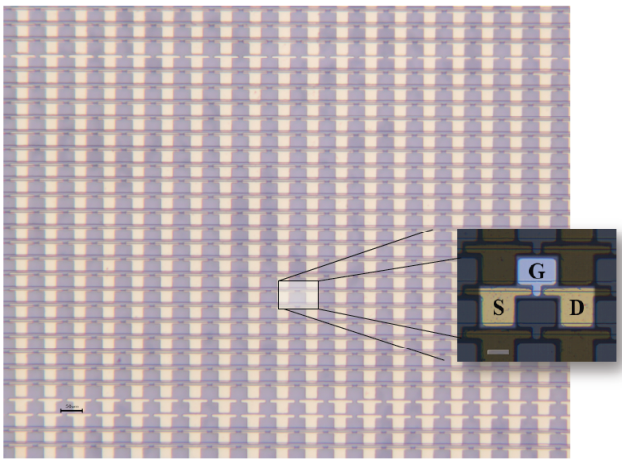

## Beyond the extended sheet: structured graphene growth

Besides fundamental graphene studies, actually very few applications require extended graphene sheets. Most require patterned structures and many require extremely fine structures. The predominant paradigm is to produce a large graphene sheet that is subsequently cut to the required shape, typically by using standard microelectronics lithography methods. This method is not unlike the way a tailor would produce a garment from a sheet of material.

*Fig. 7*
Optical image of a portion of a 10,000 array of transistors produced using sidewall graphitized channels. The inset shows one of the transistors: (S: source, D: drain, G: gate). Scale bar is 10 µm.

However other methods may well be more efficient, where the graphene is grown in its ultimate form, much like a knitted garment is produced. The advantages are clear: this method avoids damaging the graphene that certainly degrades its transport properties. The GIT group developed several approaches along these lines. The most promising of these relies on the graphitization of structured silicon carbide surfaces.

Our earlier experiments demonstrated that various crystal faces of silicon carbide graphitize at different rates. For example, at a given temperature, the Si face graphitizes much slower than the C face. We conjectured that the sidewall of a step etched into the C face of a SiC crystal would graphitize much more rapidly than the Si face so that using correct temperatures and annealing times, the sidewalls would graphitize while the horizontal surfaces would remain ungraphitized.

This guess turned out to be correct[30]. We recently demonstrated that very narrow graphene ribbons can be produced by graphitizing either natural steps on the surface, or steps that had been produced using lithography and etching methods (Fig. 7). While this route has only been begun to be explored the initial results are quite promising, and the graphene ribbons made by this method seem to have remarkably high mobilities.

**Conclusion**

This very brief overview summarizes the pioneering phases of the development of graphene based electronics at the Georgia Institute of Technology. The epitaxial graphene on silicon carbide path to graphene based electronics, pioneered by the GIT group and developed by its immediate collaborators world wide, has been adopted by many research groups around the world. Seyller and co-workers developed an inert gas annealing process to produce high quality graphene on the Si face while Starke's group found that the buffer layer on the Si face could be turned in to a graphene layer by hydrogen passivation. A large DARPA sponsored effort is currently focused on epitaxial graphene on SiC and recently 100 GHz transistors have been produced by the IBM group funded by this project using the methods developed by us[32]. A collaborative project in Sweden and England has recently demonstrated that epitaxial graphene Hall bars are ultrahigh precision resistance standards, using the quantum Hall effect at liquid helium temperature.

While we are still very far from our ultimate goal of producing integrated epitaxial graphene circuits, there are no obvious show-stopping impediments towards this goal. It is likely that this development will progress slowly but ultimate success seems to be very likely, and it is simply a matter of time before the first such devices will be commercially available.

In closing, I would like to address a major misconception that the ultimate goal of graphene based electronics is to replace silicon. This is not true. It is not likely that silicon will ever be replaced. It has established its role in electronics probably for centuries to come. However the point is not to replace silicon but to find a successor for it. That is, to develop a material that can outperform silicon where silicon hits the "brick wall", which is in speed, feature size and energy consumption. However, for applications where those limitations are not a concern, silicon will continue to be the material of choice because it is not likely that any form of electronics will be cheaper that silicon in the foreseeable future.

The appropriate analogy is probably with transportation. Goods are carried both by planes and by boats. The metal hulled ship of today is essentially identical the ship built a century ago. Shipbuilding has matured and there is little innovation left. Sea transport is slow but cheap. Planes on the other hand, are fast and the technology is still significantly advancing after the first pioneering flights a century ago. Air travel is fast but expensive. But both modes of transportation coexist harmoniously. So it might eventually be with silicon and graphene based electronics. The way I see it, epitaxial graphene electronics has just made the equivalent of its first 100 foot flight, and has still a long way to go before it will be able to make its full impact in electronics.

**References**

1. Acheson, E. Deflocculated graphite. *Journal Of The Franklin Institute* **164**, 0375-0382 (1907).

2. Boehm, H., Clauss, A., Hofmann, U & Fischer, G. Dunnste Kohlenstof-Folien. *Z Fur Naturforschung* **B 17**, 150-& (1962).

3. Boehm, H., Setton, R. & Stumpp, E. Nomenclature and terminology of graphite intercalation compounds. *Carbon* **24**, 241-245 (1986).

4. van Bommel, A.J., Crombeen, J.E. & van Tooren, A. LEED and Auger-electron Observations of the SIC(0001) Surface. *Surface Science* **48**, 463-472 (1975).

5. Forbeaux, I., Themlin, J.M. & Debever, J.M. Heteroepitaxial graphite on 6H-SiC(0001): Interface formation through conduction-band electronic structure. *Physical Review B: Condensed Matter and Materials Physics* **58**, 16396-16406 (1998).

6. Gall, N.R., RutKov, E.V. & Tontegode, A.Y. Two dimensional graphite films on metals and their intercalation. *International Journal of Modern Physics B* **11**, 1865-1911 (1997).

7. Lu, X., Yu, M., Huang, H. & Ruoff, R. Tailoring graphite with the goal of achieving single sheets. *Nanotechnology* **10**, 269-272 (1999).

8. Wallace, P.R. The Band Theory of Graphite. *Physical Review* **71**, 622-634 (1947).

9. McClure, J.W. Band Structure of Graphite and de Haas-van Alphen Effect. *Physical Review* **108**, 612-618 (1957).

10. Ando, T., Nakanishi, T. & Saito, R. Berry's phase and absence of back scattering in carbon nanotubes. *Journal of the Physical Society of Japan* **67**, 2857-2862 (1998).

11. Iijima, S. Helical Microtubules Of Graphitic Carbon. *Nature* **354**, 56-58 (1991).

12. Mintmire, J., Dunlap, B. & White, C. Are Fullerene Tubules Metallic. *Physical Review LetterS* **68**, 631-634 (1992).

13. Frank, S., Poncharal, P., Wang, Z.L. & Heer, W.A. Carbon Nanotube Quantum Resistors. *Science* **280**, 1744-1746 (1998).

14. White, C. & Todorov, T. Carbon nanotubes as long ballistic conductors. *Nature* **393**, 240-242 (1998).

15. Tans, S., Verschueren, R. & Dekker, C. Room temperature transistor based on a single carbon nanotube. *Nature* **393**, 49-52 (1998).

16. Nakada, K., Fujita, M., Dresselhaus, G. & Dresselhaus, M.S. Edge state in graphene ribbons: Nanometer size effect and edge shape dependence. *Physical Review B: Condensed Matter and Materials Physics* **54**, 17954-17961 (1996).

17.	de Heer, W. Early Development of Graphene Electronics. (2009).at <http://hdl.handle.net/1853/31270>

18.	Novoselov, K.S. et al. Electric Field Effect in Atomically Thin Carbon Films. *Science* **306**, 666-669 (2004).

19.	Berger, C. et al. Ultrathin epitaxial graphite: 2D electron gas properties and a route toward graphene-based nanoelectronics. *Journal of Physical Chemistry B* **108**, 19912-19916 (2004).

20.	Berger, C. et al. Evidence for 2D electron gas behavior in ultrathin epitaxial graphite on a SiC substrate. *Bull. Amer. Phys. Soc.* (2004).at <http://flux.aps.org/meetings/YR04/MAR04/baps/abs/S170008.html>

21.	Wu, X. et al. Half integer quantum Hall effect in high mobility single layer epitaxial graphene. *Applied Physics Letters* **95**, 223108 (2009).

22.	Berger, C. et al. Electronic Confinement and Coherence in Patterned Epitaxial Graphene. *Science* **312**, 1191-1196 (2006).

23.	Sadowski, M.L., Martinez, G., Potemski, M., Berger, C. & de Heer, W.A. Landau Level Spectroscopy of Ultrathin Graphite Layers. *Physical Review Letters* **97**, 266405 (2006).

24.	Orlita, M. et al. Approaching the Dirac Point in High-Mobility Multilayer Epitaxial Graphene. *Physical Review Letters* **101**, 267601-4 (2008).

25.	Hass, J. et al. Why multilayer graphene on 4H-SiC(000$1$) behaves like a single sheet of graphene. *Physical Review Letters* **100**, 125504 (2008).

26.	Sprinkle, M. et al. First direct observation of a nearly ideal graphene band structure. *Phys. Rev. Lett.* **103**, 4 (2009).

27.	Miller, D.L. et al. Observing the Quantization of Zero Mass Carriers in Graphene. *Science* **324**, 924-927 (2009).

28.	Sun, D. et al. Ultrafast Relaxation of Excited Dirac Fermions in Epitaxial Graphene Using Optical Differential Transmission Spectroscopy. *Physical Review Letters* **101**, 157402-4 (2008).

29.	Wu, X. et al. Epitaxial-graphene/graphene-oxide junction: An essential step towards epitaxial graphene electronics. *Physical Review Letters* **101**, 026801 (2008).

30.	Bekyarova, E. et al. Chemical Modification of Epitaxial Graphene: Spontaneous Grafting of Aryl Groups. *Journal of the American Chemical Society* **131**, 1336-1337 (2009).

31.	Sprinkle, M. et al. Scalable templated growth of graphene nanoribbons on SiC. *Nature Nanotechnology* **5**, 727-731 (2010).

32.	Kedzierski, J. et al. Epitaxial graphene transistors on SiC substrates. *IEEE Transactions on Electron Devices* **55**, 2078-2085 (2008).

33. K.V. Emtsev, A. Bostwick, K. Horn, et al. Towards wafer-size graphene layers by atmospheric pressure graphitization of SiC(0001) *Nature Materials* **8** 203 (2009).